\documentclass{iopart}
\usepackage{iopams}

\newtheorem{theorem}{Theorem}[section]
\newtheorem{proposition}[theorem]{Proposition}

\newtheorem{remark}[theorem]{Remark}
\eqnobysec
\begin{document}
\title[Hankel determinant formula for the Toda equation]{A Remark on the Hankel Determinant Formula for Solutions of the Toda Equation}

\author{Kenji Kajiwara$^1$, Marta Mazzocco$^2$ and Yasuhiro Ohta $^3$}

\address{$^1$ Graduate School of Mathematics, Kyushu University, 6-10-1 Hakozaki, Fukuoka 812-8581, Japan}
\address{$^2$ School of Mathematics, The University of Manchester,
Sackville Street, Manchester M60 1QD, United Kingdom}
\address{$^3$ Department of Mathematics, Kobe University, Rokko, Kobe 657-8501, Japan}

\begin{abstract}
We consider the Hankel determinant formula of the $\tau$ functions of
the Toda equation. We present a relationship between the determinant
formula and the auxiliary linear problem, which is characterized by a
compact formula for the $\tau$ functions in the framework of the KP
theory.  Similar phenomena that have been observed for the Painlev\'e II
and IV equations are recovered. The case of finite lattice is also
discussed.
\end{abstract}



\maketitle

\section{Introduction}
The Toda equation\cite{Toda}
\begin{equation}
\frac{d^2y_n}{dt^2}={\rm e}^{y_{n-1}-y_n}-{\rm e}^{y_{n}-y_{n+1}},\label{Toda:y}
\end{equation}
where $n\in\mathbb{Z}$, is one of the most important integrable
systems. It can be expressed in various forms such as
\begin{eqnarray}
&& \frac{dV_n}{dt}=V_n(I_n-I_{n+1}),\quad 
\frac{dI_n}{dt}=V_{n-1}-V_{n},\label{Toda:I-V}\\
&&
\frac{d\alpha_n}{dt}=\alpha_n(\beta_{n+1}-\beta_n),\quad
\frac{d\beta_n}{dt}=2(\alpha_n^2-\alpha_{n-1}^2),\label{Toda:Flaschka}
\end{eqnarray}
where the dependent variables are related to $y_n$ as
\begin{equation}
V_n = {\rm e}^{y_n - y_{n+1}},\quad
I_n = \frac{dy_n}{dt},\quad
\alpha_n=\frac{1}{2}{\rm e}^{\frac{y_n-y_{n+1}}{2}},\quad 
\beta_n=-\frac{1}{2} \frac{dy_n}{dt}. \label{Toda:vars}
\end{equation}
The Toda equation can be reduced to the bilinear equation
\begin{equation}
 \tau_n''\tau_n - (\tau_n')^2 = \tau_{n+1}\tau_{n-1} ,\label{Toda:bl}
\end{equation}
by the dependent variable transformation
\begin{equation}
y_n=\log\frac{\tau_{n-1}}{\tau_n},\quad
V_n=\frac{\tau_{n+1}\tau_{n-1}}{\tau_n^2},\quad
I_n=\frac{d}{dt}\log\frac{\tau_{n-1}}{\tau_n}. \label{Toda:vars_transform}
\end{equation}

In general, the determinant structure of the $\tau$ function (dependent
variable of bilinear equation) is the characteristic property of integrable
systems. For example, the Casorati determinant formula of the N-soliton
solution of the Toda equation (see, for example, \cite{Ueno-Takasaki,Hirota-Ito-Kako:2dTL})
\begin{equation}
\fl \tau_n = e^{\frac{t^2}{2}}\det(f_{n+j-1}^{(i)})_{i,j=1,\ldots,N},\quad
f_n^{(k)}=p_k^n~e^{p_kt+\eta_{k0}} + p_k^{-n}~e^{\frac{1}{p_k}t+\xi_{k0}},
\end{equation}
where $p_k$, $\eta_{k0}$ and $\xi_{k0}$ ($k=1,\ldots,N$) are constants,
is a direct consequence of the Sato theory; the solution space of
soliton equations is the universal Grassman manifold, on which infinite
dimensional Lie algebras are acting \cite{Jimbo-Miwa:RIMS,Miwa-Jimbo-Date:Soliton,Ueno-Takasaki}. 

If we consider the Toda equation on semi-infinite or finite lattice, the
soliton solutions do not exist but another determinantal solution
arises. For the semi-infinite case we impose the boundary condition as
\begin{equation}
\tau_{-1}=0,\quad \tau_0=1,\quad V_{0}=0,\quad n\geq 0.
\end{equation}
Then $\tau_n$ admits the Hankel determinant formula\cite{Leznov-Saveliev,Hirota:2dTM,Hirota:d2dTM}
\begin{equation}
\tau_n = \det(a_{i+j-2})_{i,j=1,\ldots,n},\quad a_0 = \tau_1,\quad a_i =
 a_{i-1}',\quad n\in\mathbb{Z}_{\geq 0}. \label{Darboux}
\end{equation}
The important feature of this determinant formula is that the lattice
site $n$ appears as the determinant size, while for the soliton
solutions the determinant size describes the number of solitons.  This
type of determinant formula is actually a special case of the
determinant forula for the infinite lattice\cite{KMNOY:Toda}. However
the meaning of the formula has not been yet fully understood.

The purpose of this article is to establish a characterization of the
Hankel determinant formula of the Toda equation; entries of the matrices
in the determinant formula are closely related to the solution of
auxiliary linear problem. Moreover, this relationship can be described
by a compact formula in the framework of the theory of KP hierarchy.

In section 2, we discuss the Hankel determinant formula of the infinite
Toda equation and present the relationship between the determinant
formula and auxiliary linear problem. In section 3, we apply the results
to the Painlev\'e II equation. We consider the case of finite lattice in
section 4.

\section{Hankel determinant formula of the solution of the Toda
 equation}
\subsection{Determinant formula and auxiliary linear problem}\label{sec:toda}

The Hankel determinant formula for $\tau_n$ satisfying the infinite Toda equation
\eref{Toda:bl} is given by as follows:
\begin{proposition}\label{prop:det}\cite{KMNOY:Toda} 
For fixed $k\in\mathbb{Z}$, we have:
\begin{equation}\label{Toda:det}
 \frac{\tau_{k+n}}{\tau_k}=\left\{
\begin{array}{ll}
 \det(a_{i+j-2}^{(k)})_{i,j=1,\ldots,n} & n>0,\\
 1 &  n=0,\\
 \det(b_{i+j-2}^{(k)})_{i,j=1,\ldots,|n|} & n<0,
\end{array}
\right.
\end{equation}
\begin{equation}\label{Toda:rec}
 \left\{
\begin{array}{l}\medskip
{\displaystyle  a_i^{(k)} = a_{i-1}^{(k)\prime} + \frac{\tau_{k-1}}{\tau_k}\sum_{l=0}^{i-2}
  a_l^{(k)}a_{i-2-l}^{(k)},\quad a_0^{(k)}=\frac{\tau_{k+1}}{\tau_k},}\\
{\displaystyle b_i^{(k)} = b_{i-1}^{(k)\prime} + \frac{\tau_{k+1}}{\tau_k}\sum_{l=0}^{i-2}
  b_l^{(k)}b_{i-2-l}^{(k)}, \quad b_0^{(k)}=\frac{\tau_{k-1}}{\tau_k}.}
\end{array}\right.
\end{equation}
\end{proposition}
We shall now relate the determinant formula to the auxiliary linear
problem of the Toda equation \eref{Toda:I-V} given by
\begin{equation}\label{Toda:lin_I-V}
 \left\{
\begin{array}{l}\medskip
{\displaystyle V_{n-1}\Psi_{n-1} + I_n\Psi_n + \Psi_{n+1}=\lambda  \Psi_n}, \\
{\displaystyle \frac{d\Psi_n}{dt}=V_{n-1}\Psi_{n-1}},
\end{array}\right. 
\end{equation}
or 
\begin{equation}
 \left\{
\begin{array}{l}\medskip
{\displaystyle L_n\Psi_n=\lambda \Psi_n,} \\
{\displaystyle \frac{d\Psi_n}{dt}=B_n\Psi_{n},}
\end{array}\right.\label{Toda:lin_I-V2}
\end{equation}
where
\begin{equation}
 L_n=V_{n-1}e^{-\partial_{n}} + I_n + e^{\partial_n},\quad B_n=V_{n-1}e^{-\partial_n}.\label{Toda:lin_I-V-LB}
\end{equation}
The adjoint linear problem associated with the linear problem 
\eref{Toda:lin_I-V} is given by
\begin{equation}
\left\{
\begin{array}{l}
{\displaystyle \Psi^*_{n-1} + I_n\Psi^*_n + V_n\Psi^*_{n+1}=\lambda \Psi^*_n,}\\[2mm]
{\displaystyle \frac{d\Psi^*_n}{dt}=-V_{n}\Psi^*_{n+1},}
\end{array}
\right.
\label{Toda:adjlin_I-V}
\end{equation}
or
\begin{equation}
 \left\{
\begin{array}{l}\medskip
{\displaystyle L_n^*\Psi^*_n=\lambda \Psi^*_n,} \\
{\displaystyle -\frac{d\Psi^*_n}{dt}=B_n^*\Psi^*_{n},}
\end{array}\right.\label{Toda:adjlin_I-V2}
\end{equation}
where
\begin{equation}
 L_n^*=V_{n}e^{\partial_n} + I_n + e^{-\partial_n},\quad B_n^*=V_{n}e^{\partial_n}.\label{Toda:adjlin_I-V-LB}
\end{equation}

The compatibility condition for each problem
\begin{equation}
 \frac{dL_n}{dt}=[B_n,L_n],\quad \frac{dL_n^*}{dt}=[-B_n^*,L_n^*]
\end{equation}
yields the Toda equation \eref{Toda:I-V}, respectively.

One of our main results is that the entries of the determinants in the
Hankel determinant formula arise as the coefficients of asymptoric
expansions at $\lambda=\infty$ of the ratio of solutions of the linear
and adjoint linear problems. To state the result more precisely, 
we define
\begin{equation}
 \Xi_k(t,\lambda)=\frac{\Psi_k(t,\lambda)}{\Psi_{k+1}(t,\lambda)},\quad
\Omega_k(t,\lambda)=\frac{\Psi^*_{k+1}(t,\lambda)}{\Psi^*_{k}(t,\lambda)}.\label{Toda:Xi_and_Omega}
\end{equation}
\begin{theorem}\label{thm:Toda1}
\begin{enumerate}
 \item The ratios $\Xi_k(t,\lambda)$ and $\Omega_k(t,\lambda)$ admits two kinds of
       asymptotic expansions as functions of $\lambda$ as $\lambda\to\infty$
\begin{eqnarray}
 \Xi_k^{(-1)}(t,\lambda)&=&u_{-1}\lambda^{-1}+u_{-2}\lambda^{-2}+\cdots,\label{Toda:Xi_expansion1}\\
 \Xi_k^{(1)}(t,\lambda) &=&v_{1}\lambda+v_{0}+ v_{-1}\lambda^{-1}+\cdots,\label{Toda:Xi_expansion-1}
\end{eqnarray}
and
\begin{eqnarray}
 \Omega_k^{(-1)}(t,\lambda)&=&u_{-1}\lambda^{-1}+u_{-2}\lambda^{-2}+\cdots,\label{Toda:Omega_expansion1}\\
 \Omega_k^{(1)}(t,\lambda) &=&v_{1}\lambda+v_{0}+ v_{-1}\lambda^{-1}+\cdots,\label{Toda:Omega_expansion2}
\end{eqnarray}
respectively.
 \item The above asymptotic expantions are related to the Hankel
       determinants entries $a_i^{(k)}$ and $b_i^{(k)}$ as follows:
 \begin{eqnarray}
\fl &&   \Xi_k^{(-1)}(t,\lambda)=\frac{1}{\lambda}\frac{\tau_k}{\tau_{k-1}}
\sum_{i=0}^\infty b_i^{(k)}~\lambda^{-i},\label{Xi:1}\\
\fl && \Xi_k^{(1)}(t,\lambda)=\frac{\tau_k^2}{\tau_{k+1}\tau_{k-1}}
\left[
\lambda -
\frac{\left(\frac{\tau_k}{\tau_{k+1}}\right)'}{\frac{\tau_k}{\tau_{k+1}}}
-\frac{1}{\lambda}\frac{\tau_{k}}{\tau_{k+1}}\sum_{i=0}^\infty a_i^{(k+1)}~(-\lambda)^{-i}
\right],\label{Xi:2}
 \end{eqnarray}
and
 \begin{eqnarray}
\fl &&   \Omega_k^{(-1)}(t,\lambda)=\frac{1}{\lambda}\frac{\tau_k}{\tau_{k+1}}
\sum_{i=0}^\infty a_i^{(k)}~(-\lambda)^{-i},\label{Omega:1}\\
\fl && \Omega_k^{(1)}(t,\lambda)=\frac{\tau_k^2}{\tau_{k+1}\tau_{k-1}}
\left[
\lambda - \frac{\left(\frac{\tau_{k-1}}{\tau_{k}}\right)'}{\frac{\tau_{k-1}}{\tau_{k}}}
-\frac{1}{\lambda}\frac{\tau_{k}}{\tau_{k-1}}\sum_{i=0}^\infty b_i^{(k-1)}~\lambda^{-i}
\right].\label{Omega:2}
\end{eqnarray}
 \item $\Xi_k^{(\pm 1)}$ and $\Omega_k^{(\pm 1)}$ are related as follows:
\begin{equation}\fl
\Omega_k^{(1)}(t,\lambda)~\Xi_k^{(-1)}(t,\lambda)=\frac{\tau_k^2}{\tau_{k+1}\tau_{k-1}},\quad
\Omega_k^{(-1)}(t,\lambda)~\Xi_k^{(1)}(t,\lambda)=\frac{\tau_k^2}{\tau_{k+1}\tau_{k-1}}.
\label{Toda:Omega_and_Xi}
\end{equation}
\end{enumerate}
\end{theorem}
\textbf{Brief sketch of the proof of Theorem \ref{thm:Toda1}.} 
One can prove Theorem \ref{thm:Toda1} by direct calculation. From the linear problem \eref{Toda:lin_I-V} and
\eref{Toda:Xi_and_Omega}, we see that $\Xi_k(t,\lambda)$ satisfies the Riccati equation
\begin{equation}
 \frac{\partial \Xi_k}{\partial t}=-V_{k}\Xi_k^2 +
 (\lambda-I_{k})\Xi_k -1.\label{Riccati:Xi}
\end{equation}
Plugging series expansion $\Xi_k=\lambda^\rho\sum_{i=0}^\infty
h_i~\lambda^{-i}$ into \eref{Riccati:Xi} and considering the balance of
leading terms, we find that $\rho$ must be $\rho=1,-1$, which proves
\eref{Toda:Xi_expansion1} and \eref{Toda:Xi_expansion-1} . Moreover, it
is possible to verify \eref{Xi:1} and \eref{Xi:2} by deriving recursion
relations of coefficients for each case and comparing them with
\eref{Toda:rec}. Similarly, from the Riccati equation for $\Omega$
\begin{equation}
  \frac{\partial \Omega_k}{\partial t}=V_{k}\Omega_k^2 +
 (-\lambda+I_{k+1})\Omega_k +1, \label{Riccati:Omega}
\end{equation}
one can prove the statements for $\Omega$. For (iii), putting
$X_k=\frac{\tau_k^2}{\tau_{k+1}\tau_{k-1}}\frac{1}{\Xi_k^{(-1)}}
=\frac{1}{V_k\Xi_k^{(-1)}}$, plugging this expression into the Riccati
equation \eref{Riccati:Omega} and using \eref{Toda:I-V}, we find that
$X_k$ satisfies \eref{Riccati:Xi}. Since the expansion of $\Xi_k^{(-1)}$
starts from $\lambda^{-1}$, the leading order of $X_k$ is $\lambda$ and
thus $X_k=\Omega^{(1)}$. The second equation of \eref{Toda:Omega_and_Xi}
can be proved in a similar manner. $\square$

\subsection{KP theory}
The results in the previous section can be characterized by a compact
formula in terms of the language of the KP
theory\cite{Jimbo-Miwa:RIMS,Miwa-Jimbo-Date:Soliton,OSTT:Sato}.

We introduce infinitely many independent variables
$x=(x_1,x_2,x_3,\ldots)$, $x_1=t$, and let $\tau_n(x)$ be the $\tau$
function of the one-dimensional Toda lattice
hierarchy\cite{Ueno-Takasaki,Jimbo-Miwa:RIMS} and the first modified KP
hierarchy\cite{Jimbo-Miwa:RIMS}. Namely, $\tau_n$, $n\mathbb{Z}$,
satisfy the following bilinear equations
\begin{eqnarray}
\fl &&
 D_{x_1}p_{j+1}\left(\frac{1}{2}\tilde{D}\right)\tau_{n}\cdot\tau_{n}
=p_j\left(\frac{1}{2}\tilde D\right)\tau_{n+1}\cdot\tau_{n-1},\quad
j=0,1,2\ldots, \label{bl:Toda_hierarchy}
\\
\fl && \left[
D_{x_1}p_j\left(\frac{1}{2}\tilde  D\right)-p_{j+1}\left(\frac{1}{2}\tilde D\right)
+p_{j+1}\left(-\frac{1}{2}\tilde D\right)
\right]\tau_{n+1}\cdot\tau_{n}=0,\quad j=0,1,2\ldots,\label{bl:1st_modified_KP_hierarchy}
\end{eqnarray}
where $p_0(x)$, $p_1(x)$, $\cdots$ are the elementary Schur functions
\begin{equation}
 \sum_{n=0}^\infty p_n(x)\kappa^n = \exp\sum_{i=1}^\infty x_i \kappa^i,\label{elementary_Schur}
\end{equation}
and
\begin{equation}
 \tilde D = (D_{x_1}, \frac{1}{2}D_{x_2},\ldots,\frac{1}{n}D_{x_n},\ldots),
\end{equation}
$D_{x_i}$ $(i=1,2,\ldots)$ being the Hirota's $D$-operator. Then we
have the following formula:
\begin{proposition}\label{prop:KP}
For fixed $k\in\mathbb{Z}$, we have
\begin{equation}\label{KP:det}
 \frac{\tau_{k+n}}{\tau_k}=\left\{
\begin{array}{ll}
 \det(a_{i+j-2}^{(k)})_{i,j=1,\ldots,n} & n>0,\\
 1 &  n=0,\\
 \det(b_{i+j-2}^{(k)})_{i,j=1,\ldots,|n|} & n<0,
\end{array}
\right.
\end{equation}
where
\begin{equation}\label{KP:tau}
 a_i^{(k)}=p_i(\tilde\partial)\frac{\tau_{k+1}}{\tau_k},\quad
 b_i^{(k)}=(-1)^ip_i(-\tilde\partial)\frac{\tau_{k-1}}{\tau_k},
\end{equation}
and
\begin{equation}
 \tilde\partial=(\partial_{x_1}, \frac{1}{2}\partial_{x_2},\ldots,\frac{1}{n}\partial_{x_n},\ldots).
\end{equation}
\end{proposition}
\begin{remark}
It might be interesting to remark here that
$a_0^{(k)}=\frac{\tau_{k+1}}{\tau_k}$ and
$b_0^{(k)}=\frac{\tau_{k-1}}{\tau_k}$ satisfy the nonlinear
Schr\"odinger hierarchy. In fact, Equations \eref{Toda:rec} and \eref{KP:tau} with
$i=2$ imply for $a=a_0^{(k)}$ and $b=b_0^{(k)}$
\begin{equation}
 a_{x_2}= a_{x_1x_1} + 2a^2b,\quad b_{x_2} = - (b_{x_1x_1}+2a^2b).
\end{equation}
Similarly, for $i=3$ we have
\begin{equation}
 a_{x_3}=a_{x_1x_1x_1} + 6aba_{x_1},\quad
 b_{x_3}=b_{x_1x_1x_1} + 6abb_{x_1}.
\end{equation}
\end{remark}
Before proceeding to the proof, we note that the auxiliary linear
problem \eref{Toda:lin_I-V} and its adjoint problem
\eref{Toda:adjlin_I-V} are recovered from the bilinear equations
\eref{bl:Toda_hierarchy} and \eref{bl:1st_modified_KP_hierarchy}.
In fact, suppose that $\tau_n$ depends on a discrete independent variable $l$ and
satisfies the discrete modified KP equation
\begin{eqnarray}
&& D_{x_1}\tau_n(l+1)\cdot\tau_n(l)=-\frac{1}{\lambda}~\tau_{n+1}(l+1)\tau_{n-1}(l),
\label{bl:discrete_modified_KP1}
\\
&&(\frac{1}{\lambda} D_{x_1}+1)\tau_{n+1}(l+1)\cdot\tau_n(l)-\tau_n(l+1)\tau_{n+1}(l)=0,
\label{bl:discrete_modified_KP2}
\end{eqnarray}
then one can show that Equations \eref{bl:Toda_hierarchy} and
\eref{bl:1st_modified_KP_hierarchy} are equivalent to
\eref{bl:discrete_modified_KP1} and
\eref{bl:discrete_modified_KP2}, respectively, through the Miwa
transformation\cite{Miwa:dKP,Jimbo-Miwa:RIMS}
\begin{equation}\fl
 x_n = \frac{l}{n(-\lambda)^n}\quad\mbox{or}\quad 
\frac{\partial}{\partial l}
=-\frac{1}{\lambda}\frac{\partial}{\partial x_1}
+\frac{1}{2\lambda^2}\frac{\partial}{\partial x_2}+\cdots
+\frac{1}{j(-\lambda)^j}\frac{\partial}{\partial x_j}+\cdots.\label{Miwa_transform}
\end{equation}
Putting 
\begin{equation}
 \Psi^*_n = \lambda^{-n} \frac{\tau_n(l+1)}{\tau_n(l)},\label{Psi*_and_tau}
\end{equation}
\begin{equation}
 V_n = \frac{\tau_{n+1}(l)\tau_{n-1}(l)}{\tau_n(l)^2},\quad
I_n = \frac{d}{dt}\log\frac{\tau_{n-1}(l)}{\tau_n(l)},\label{I-V_and_tau}
\end{equation}
and noticing $t=x_1$, the bilinear equations \eref{bl:discrete_modified_KP1} and
\eref{bl:discrete_modified_KP2} are rewritten as
\begin{equation}
\begin{array}{l}
\medskip
 {\displaystyle \Psi^{*\prime}_n=-V_{n+1}\Psi^*_{n+1},}\\
{\displaystyle \Psi^*_{n}+I_{n+1}\Psi^*_{n+1}+V_{n+2}\Psi^*_{n+2}=\lambda\Psi^*_{n+1},}
\end{array}
\end{equation}
which are equivalent to the adjoint linear problem \eref{Toda:adjlin_I-V}.
Similarly, shifting $l\to l-1$ in \eref{bl:discrete_modified_KP1} and
\eref{bl:discrete_modified_KP2} and putting
\begin{equation}
 \Psi_{n+1} = \lambda^{n} \frac{\tau_n(l-1)}{\tau_n(l)},\label{Psi_and_tau}
\end{equation}
we obtain
\begin{equation}
\begin{array}{l}
\medskip
{\displaystyle \Psi_{n+1}'=V_n\Psi_{n},} \\
{\displaystyle V_n\Psi_{n}+I_{n+1}\Psi_{n+1}+\Psi_{n+2}=\lambda\Psi_{n+1},}
\end{array}
\end{equation}
which is also equivalent to the linear problem \eref{Toda:lin_I-V}.\\

\noindent\textbf{Proof of Proposition \ref{prop:KP}.}
From \eref{Psi_and_tau}, \eref{Miwa_transform} and \eref{elementary_Schur} we have
\begin{eqnarray*}
\fl &&\frac{\Psi_k(t,\lambda)}{\Psi_{k+1}(t,\lambda)}
=\frac{1}{\lambda}\frac{\tau_{k-1}(l-1)\tau_{k}(l)}{\tau_{k}(l-1)\tau_{k-1}(l)}
=\frac{1}{\lambda}\frac{\tau_k(l)}{\tau_{k-1}(l)}~
{\rm e}^{-\frac{\partial}{\partial l}}\frac{\tau_{k-1}(l)}{\tau_{k}(l)}\\
\fl &=&\frac{1}{\lambda}\frac{\tau_k(l)}{\tau_{k-1}(l)}~{\rm e}^{-\sum\limits_{j=1}^\infty
\frac{1}{j(-\lambda)^j}\frac{\partial}{\partial x_j}}\frac{\tau_{k-1}(l)}{\tau_{k}(l)}
=
\frac{1}{\lambda}\frac{\tau_k(l)}{\tau_{k-1}(l)}~
\sum_{n=0}^\infty p_n(-\tilde\partial)\frac{\tau_{k-1}(l)}{\tau_{k}(l)}~(-\lambda)^{-n}.
\end{eqnarray*}
Therefore \Eref{Xi:1} in Theorem \ref{thm:Toda1} implies
\begin{equation}
b_n^{(k)}=(-1)^np_n(-\tilde\partial)\frac{\tau_{k}(l)}{\tau_{k-1}(l)}.
\end{equation}
Similarly we have from \eref{Psi*_and_tau}, \eref{Miwa_transform} and \eref{elementary_Schur}
\begin{eqnarray*}
\fl &&\frac{\Psi^*_{k+1}(t,\lambda)}{\Psi^*_{k}(t,\lambda)}
=\frac{1}{\lambda}\frac{\tau_{k+1}(l+1)}{\tau_k(l+1)}\frac{\tau_k(l)}{\tau_{k+1}(l)}
=\frac{1}{\lambda}\frac{\tau_k(l)}{\tau_{k+1}(l)}
{\rm e}^{\frac{\partial}{\partial l}}\frac{\tau_{k+1}(l)}{\tau_{k}(l)}\\
\fl &=&\frac{1}{\lambda}\frac{\tau_k(l)}{\tau_{k+1}(l)}~
{\rm e}^{\sum\limits_{j=1}^\infty
\frac{1}{j(-\lambda)^j}\frac{\partial}{\partial x_j}}
\frac{\tau_{k+1}(l)}{\tau_{k}(l)}
=\frac{1}{\lambda}\frac{\tau_k(l)}{\tau_{k+1}(l)}~
\sum_{n=0}^\infty p_n(\tilde\partial)\frac{\tau_{k+1}(l)}{\tau_{k}(l)}(-\lambda)^{-n}.
\end{eqnarray*}
Therefore comparing with \eref{Omega:1} we obtain
\begin{equation}
a_n^{(k)}=p_n(\tilde\partial)\frac{\tau_{k+1}(l)}{\tau_{k}(l)},
\end{equation}
which proves Proposition \ref{prop:KP}. $\square$

\section{Painlev\'e equations}
\subsection{Local Lax pair}
Originally the relations between determinant formula of the solutions
and auxiliary linear problem have been derived for the Painlev\'e II and
IV equations \cite{JKM:p2,JKM:p4}. In the particular case of the
rational solutions of the Painlev\'e II and IV equations these results
give the relation between determinant formula and the Airy function found in
\cite{IKN:p2,GK:p4}. It may be natural to regard those relationships as
originating from the Toda equation, since the sequence of $\tau$
functions generated by the B\"acklund transformations of Painlev\'e
equations are described by the Toda equation
\cite{Okamoto:p6,Okamoto:p5,Okamoto:p24,Okamoto:p3,JM:Monodromy,KMNOY:Toda}.
In this section, we show that the results for the Painlev\'e II equation
can be recovered from the results in the section 2.  The key ingredient
of the correspondence is the local Lax pair, which is the auxiliary
linear problem for Toda equation formulated by a pair of $2\times 2$
matrices\cite{Fadeev-Takhtajan-Tarasov}:
\begin{eqnarray}
\fl
&&\widetilde{L}_n\phi_n=\phi_{n+1},\quad
\widetilde{L}_n(t,\lambda)=\left(
\begin{array}{cc}
-I_n+\lambda  & -{\rm e}^{-y_n}\\ {\rm e}^{y_n} & 0  
\end{array}
\right),\label{local_Lax:L}\\
\fl
&&\frac{d\phi_n}{dt}=\widetilde{B}_n\phi_n,
\quad
\widetilde{B}_n(t,\lambda)=
\left(
\begin{array}{cc}
-\frac{1}{2} & 0\\
0 & \frac{1}{2}
\end{array}
\right)\lambda
+ \left(
\begin{array}{cc}
0  & {\rm e}^{-y_n}\\
-{\rm e}^{y_{n-1}} &0
\end{array}\right),
\label{local_Lax:B}
\end{eqnarray}
\begin{equation}
\phi_n=\left(\begin{array}{c}\phi_n^{(1)}  \\\phi_n^{(2)} \end{array}\right),\quad
y_n=\log\frac{\tau_{n-1}}{\tau_n}.
\end{equation}
Similarly, the adjoint linear problem is given by
\begin{eqnarray}
\fl
&& \widetilde{L}_n^*\phi^*_n=\phi^*_{n-1},\quad
\widetilde{L}_n^*(t,\lambda)=
\left(
\begin{array}{cc}
-I_n+\lambda  & {\rm e}^{y_n}\\-{\rm e}^{-y_n}  & 0  
\end{array}
\right),\label{adjlocal_Lax:L}\\
\fl
&&  \frac{d\phi^*_n}{dt}=\widetilde{B}_n^*\phi^*_n,\quad
\widetilde{B}_n^*(t,\lambda)=
\left(
\begin{array}{cc}
\frac{1}{2} & 0\\
0 & -\frac{1}{2} \\
\end{array}
\right)\lambda
+
\left(\begin{array}{cc}  0 &{\rm e}^{y_{n}} \\
-{\rm e}^{-y_{n+1}} &0 \end{array}\right),\label{adjlocal_Lax:B}
\end{eqnarray}
\begin{equation}
\phi^*_n=\left(\begin{array}{c}\phi^{*(1)}_n  \\\phi^{*(2)}_n \end{array}\right).
\end{equation}
Compatibility condition for each problem
\begin{equation}
\frac{d\widetilde{L}_n}{dt}=\widetilde{B}_{n+1}\widetilde{L}_n-\widetilde{L}_n\widetilde{B}_n,\quad 
\frac{d\widetilde{L}_n^*}{dt}=-\widetilde{B}_{n-1}^*\widetilde{L}_n^*+\widetilde{L}_n^*\widetilde{B}_n^*,
\end{equation}
gives Toda equation (\ref{Toda:I-V}), respectively. By comparing
\eref{local_Lax:L} and \eref{local_Lax:B} with \eref{Toda:lin_I-V}¡¤
similarly by comparing \eref{adjlocal_Lax:L} and \eref{adjlocal_Lax:B}
with \eref{Toda:adjlin_I-V}, one sees that there is a relationship
between the solutions of the linear problems:
\begin{eqnarray}
&&\phi_n^{(1)}={\rm e}^{-\frac{1}{2}\lambda t}~\Psi_n,\quad
\phi_n^{(2)}={\rm e}^{-\frac{1}{2}\lambda t}\frac{\tau_{n-2}}{\tau_{n-1}}~\Psi_{n-1},\label{lin_sol:1}\\
&&\phi_n^{*(1)}={\rm e}^{\frac{1}{2}\lambda t}~\Psi^*_n,\hskip15pt
\phi_n^{*(2)}=-{\rm e}^{\frac{1}{2}\lambda t}\frac{\tau_{n+1}}{\tau_{n}}~\Psi^*_{n+1}.\label{lin_sol:2}
\end{eqnarray}
\subsection{Painlev\'e II equation}
In this section we consider the Painlev\'e II equation (P$_{\rm II}$)
\begin{equation}
 \frac{d^2u}{dt^2}=2u^3-4tu + 4\left(\alpha+\frac{1}{2}\right).\label{p2}
\end{equation}
We denote \eref{p2} as P$_{\rm II}$[$\alpha$] when it is necessary to
specify the parameter $\alpha$ explicitly. Suppose that $\tau_0$ and
$\tau_1$ satisfy the bilinear equations
\begin{eqnarray}
 && (D_t^2-2t)~\tau_1\cdot\tau_0 = 0, \label{p2:bl1}\\
 && (D_t^3-2tD_t-4(\alpha+\frac{1}{2}))~\tau_1\cdot\tau_0=0, \label{p2:bl2}
\end{eqnarray}
then it is easily verified that 
\begin{equation}
 u=\frac{d}{dt}\log\frac{\tau_1}{\tau_0}
\end{equation}
satisfies P$_{\rm II}$[$\alpha$] \eref{p2}. If we generate the sequence $\tau_n$
($n\in\mathbb{Z}$) by the Toda equation
\begin{equation}
 \frac{1}{2}D_t^2~\tau_n\cdot\tau_n = \tau_{n+1}\tau_{n-1}, \label{p2:toda}
\end{equation}
then it is shown that $\tau_n$ satisfy
\begin{eqnarray}
 && (D_t^2-2t)~\tau_{n+1}\cdot\tau_n = 0, \label{p2:bl3}\\
 && (D_t^3-2tD_t-4(\alpha+\frac{1}{2}+n))~\tau_{n+1}\cdot\tau_n=0, \label{p2:bl4}
\end{eqnarray}
and that
\begin{equation}
 u=\frac{d}{dt}\log\frac{\tau_{n+1}}{\tau_{n}}
\end{equation}
satisfies P$_{\rm II}$[$\alpha+n$].
In this sense, the Toda equation \eref{p2:toda} describes the
B\"acklund transformation of P$_{\rm II}$ (see, for example, \cite{Noumi:book}). 
Therefore one can apply Proposition \ref{prop:det} to obtain the
determinant formula: for fixed $k\in\mathbb{Z}$ we have
\begin{equation}\label{p2:det}
 \frac{\tau_{k+n}}{\tau_k}=\left\{
\begin{array}{ll}
 \det(a_{i+j-2}^{(k)})_{i,j=1,\ldots,n} & n>0,\\
 1 &  n=0,\\
 \det(b_{i+j-2}^{(k)})_{i,j=1,\ldots,|n|} & n<0,
\end{array}
\right.
\end{equation}
where 
\begin{equation}
 \left\{
\begin{array}{l}\medskip
{\displaystyle  a_i^{(k)} = a_{i-1}^{(k)\prime} + \frac{\tau_{k-1}}{\tau_k}\sum_{l=0}^{i-2}
  a^{(k)}_la^{(k)}_{i-2-l},\quad a^{(k)}_0=\frac{\tau_{k+1}}{\tau_k},}\\
{\displaystyle b^{(k)}_i = b_{i-1}^{(k)\prime} + \frac{\tau_{k+1}}{\tau_k}\sum_{l=0}^{i-2}
  b^{(k)}_lb^{(k)}_{i-2-l}, \quad b^{(k)}_0=\frac{\tau_{k-1}}{\tau_k}.}
\end{array}\right.\label{p2:rec}
\end{equation}
Now consider the auxiliary linear problem for P$_{\rm II}$[$\alpha$] \eref{p2}\cite{JM:Monodromy}:
\begin{eqnarray}
\fl
&&\frac{\partial Y}{\partial \lambda}= AY,\quad A=
\left(
\begin{array}{cc}
\frac{1}{4} & 0\\
0 & -\frac{1}{4}
\end{array}
\right)\lambda^2
+
\left(
\begin{array}{cc}
0 & -\frac{1}{2}\frac{\tau_1}{\tau_0}\\
\frac{1}{2}\frac{\tau_{-1}}{\tau_0} & 0
\end{array}
\right)\lambda \nonumber\\
\fl
&&\hskip146pt 
+ \left(
\begin{array}{cc}
-\frac{z+t}{2} & \frac{1}{2}\left(\frac{\tau_1}{\tau_0}\right)' \\[2mm]
\frac{1}{2}\left(\frac{\tau_{-1}}{\tau_0}\right)'& \frac{z+t}{2}
\end{array}
\right),\label{p2:A}
\\
\fl
&& \frac{\partial Y}{\partial t}=BY,\quad B=
\left(
\begin{array}{cc}
-\frac{1}{2} & 0\\
0 & \frac{1}{2}
\end{array}
\right)\lambda
+
\left(
\begin{array}{cc}
 0& \frac{\tau_1}{\tau_0} \\
-\frac{\tau_{-1}}{\tau_{0}} & 0
\end{array}
\right),\label{p2:B}
\end{eqnarray}
\begin{equation}
 Y=\left(\begin{array}{c}Y_1 \\Y_2\end{array}\right),\quad z=-\frac{\tau_{1}\tau_{-1}}{\tau_0^2}.\label{p2:Y}
\end{equation}
Comparing \eref{p2:B} with \eref{local_Lax:B}, we immediately find that
\begin{equation}
 B=\widetilde{B_1},\quad Y=\phi_1.
\end{equation}
We note that it is possible to regard \eref{p2:A} as the equation
defining $\lambda$-flow which is consistent with evolution in $t$. Also,
the linear equation \eref{local_Lax:L} describes the B\"acklund
transformation. Similarly, we have the adjoint problem
\begin{eqnarray}
\fl
&&\frac{\partial Y^*}{\partial \lambda}= A^*Y^*,\ A^*=
\left(
\begin{array}{cc}
\frac{1}{4} & 0\\
0 & -\frac{1}{4}
\end{array}
\right)\lambda^2
+
\left(
\begin{array}{cc}
0 & \frac{1}{2}\frac{\tau_{-1}}{\tau_0}\\[1mm]
-\frac{1}{2}\frac{\tau_1}{\tau_0} & 0
\end{array}
\right)\lambda \nonumber\\
\fl
&& \hskip159pt 
+
\left(
\begin{array}{cc}
-\frac{z+t}{2} & \frac{1}{2}\left(\frac{\tau_{-1}}{\tau_0}\right)' \\[2mm]
\frac{1}{2}\left(\frac{\tau_1}{\tau_0}\right)' & \frac{z+t}{2}
\end{array}
\right),\label{p2:A*}
\\
\fl
&&\frac{\partial Y^*}{\partial t}=B^*Y^*,\quad
B^*=
\left(
\begin{array}{cc}
\frac{1}{2} & 0\\
0 & -\frac{1}{2}
\end{array}
\right)\lambda
+
\left(
\begin{array}{cc}
 0& \frac{\tau_{-1}}{\tau_0} \\
-\frac{\tau_{1}}{\tau_0} & 0
\end{array}
\right),\quad
\label{p2:B*}
\end{eqnarray}
\begin{equation}\label{p2:Y*}
Y^*=\left(\begin{array}{c}Y_1 \\Y_2\end{array}\right),\quad z=-\frac{\tau_1\tau_{-1}}{\tau_0^2},
\end{equation}
where we have a correspondence
\begin{equation}
 B^* = \widetilde{B}^*_{0},\quad Y^*=\phi^*_{1}.
\end{equation}
Therefore, if we apply Theorem \ref{thm:Toda1} noticing \eref{lin_sol:1}
and \eref{lin_sol:2}, we have the following:
\begin{proposition}\label{prop:p2}
We put
\begin{equation}
\Lambda(t,\lambda) = \frac{Y_2}{Y_1},\quad \Pi(t,\lambda) = \frac{Y_2^*}{Y_1^*}.
\end{equation}
\begin{enumerate}
 \item The ratios $\Lambda$ and $\Pi$ admit two kinds of
       asymptotic expansions as functions of $\lambda$ as $\lambda\to\infty$
\begin{eqnarray}
 \Lambda^{(-1)}(t,\lambda)&=&u_{-1}\lambda^{-1}+u_{-2}\lambda^{-2}+\cdots,\label{p2:Lambda_expansion-1}\\
 \Lambda^{(1)}(t,\lambda) &=&v_{1}\lambda+v_{0}+ v_{-1}\lambda^{-1}+\cdots,\label{p2:Lambda_expansion1}
\end{eqnarray}
and
\begin{eqnarray}
 \Pi^{(-1)}(t,\lambda)&=&u_{-1}\lambda^{-1}+u_{-2}\lambda^{-2}+\cdots,\label{p2:Pi_expansion-1}\\
 \Pi^{(1)}(t,\lambda)&=&v_{1}\lambda+v_{0}+ v_{-1}\lambda^{-1}+\cdots,\label{p2:Pi_expansion1}
\end{eqnarray}
respectively.
 \item The above asymptotic expansions are related to the Hankel
       determinants entries $a_i^{(k)}$ and $b_i^{(k)}$ as follows:
 \begin{eqnarray}
\fl &&   \Lambda^{(-1)}(t,\lambda)=\frac{1}{\lambda}\sum_{i=0}^\infty b_i^{(0)}~\lambda^{-i},\label{Lambda:1}\\
\fl && \Lambda^{(1)}(t,\lambda)=\frac{\tau_0}{\tau_{1}}
\left[
\lambda -
\frac{\left(\frac{\tau_0}{\tau_{1}}\right)'}{\frac{\tau_0}{\tau_{1}}}
-\frac{1}{\lambda}\frac{\tau_{0}}{\tau_{1}}\sum_{i=0}^\infty a_i^{(1)}~(-\lambda)^{-i}
\right],\label{Lambda:2}
 \end{eqnarray}
and
 \begin{eqnarray}
\fl &&   \Pi^{(-1)}(t,\lambda)=\frac{1}{(-\lambda)}
\sum_{i=0}^\infty a_i^{(0)}~(-\lambda)^{-i},\label{Pi:1}\\
\fl && \Pi^{(1)}(t,\lambda)=-\frac{\tau_0}{\tau_{-1}}
\left[
\lambda - \frac{\left(\frac{\tau_{-1}}{\tau_{0}}\right)'}{\frac{\tau_{-1}}{\tau_{0}}}
-\frac{1}{\lambda}\frac{\tau_{0}}{\tau_{-1}}\sum_{i=0}^\infty b_i^{(-1)}~\lambda^{-i}
\right].\label{Pi:2}
\end{eqnarray}
 \item $\Lambda^{(\pm 1)}$ and $\Pi^{(\pm 1)}$ are related as follows:
\begin{equation}
 \fl
\Pi^{(1)}(t,\lambda)~\Lambda^{(-1)}(t,\lambda)=1,\quad
\Pi^{(-1)}(t,\lambda)~\Lambda^{(1)}(t,\lambda)=1.
\label{p2:Lambda_and_Pi}
\end{equation}
\end{enumerate}
\end{proposition}
Proposition \ref{prop:p2} is equivalent to the results presented in
\cite{IKN:p2,JKM:p2}. In other words, the relations between determinant formula
for the solution of P$_{\rm II}$ and auxiliary linear problems
originate from the structure of the Toda equation. We also note that
one can recover the results for the Painlev\'e IV equation\cite{GK:p4,JKM:p4}
in similar manner.
\section{Toda equation on finite lattice}
\subsection{Determinant formula}
Let us consider the Toda equation on the finite lattice.
Namely, we impose the boundary condition
\begin{equation}
\begin{array}{ll}
V_0=0, & V_N=0,\\
y_0=-\infty, & y_{N+1}=\infty,\\
\alpha_0=0, &  \alpha_N=0,
\end{array}
\end{equation}
on the Toda equation \eref{Toda:y},  \eref{Toda:I-V} and \eref{Toda:Flaschka}, respectively.
In order to realize this condition on the level of the $\tau$ funtion,
we proceed as follows: in the bilinear equation \eref{Toda:bl}, 
imposing the boundary condition on the left edge of lattice
\begin{equation}
\tau_{-1}=0,\quad \tau_0\neq 0,\label{finite_Toda:bc1}
\end{equation}
it immediately follows $\tau_{-2}=0$ and one can restrict the Toda
equation on the semi-infinite lattice $n\geq 0$ . In this case, the
determinant formula reduces to
\begin{equation}
 \frac{\tau_k}{\tau_0}=\det(a^{(0)}_{i+j-2})_{i,j=1,\cdots,k}\ (n\geq 1),
\quad a^{(0)}_{i+1}=a_i^{(0)\prime},\quad a_0^{(0)}=\frac{\tau_1}{\tau_0},
\label{Darboux:2}
\end{equation}
which is equivalent to \eref{Darboux}. Moreover, imposing the boundary
condition on the right edge of lattice
\begin{equation}
\tau_N\neq 0,\quad \tau_{N+1}=0, \label{finite_Toda:bc2}
\end{equation}
then we have the finite Toda equation
\begin{equation}
 \tau_n''\tau_n-\left(\tau_n'\right)^2=\tau_{n+1}\tau_{n-1},\quad
  n=0,\cdots,N,\quad \tau_{-1}=\tau_{N+1}=0.\label{finite_Toda}
\end{equation}
It is easily verified that the boundary condition \eref{finite_Toda:bc2}
is satisfied by putting 
\begin{equation}
 a^{(0)}_0=\sum_{i=1}^N c_i {\rm e}^{\mu_i t},
\end{equation}
where $c_i$ and $\mu_i$ $(i=1,\ldots,N)$ are arbitrary constants.

It is sometimes convenient to consider the finite Toda equation in the
form of \eref{Toda:Flaschka}. One reason for this is that the auxiliary
linear problem associated with \eref{Toda:Flaschka}
\begin{equation}
\fl
\alpha_{n-1}\Phi_{n-1}+\beta_n\Phi_n +
  \alpha_{n}\Phi_{n+1}=\mu\Phi_n,\quad
\frac{d\Phi_n}{dt}=-\alpha_{n-1}\Phi_{n-1}+\alpha_{n}\Phi_{n+1},
\label{lin:Flaschka}
\end{equation}
or
\begin{equation}
L\Phi=\mu\Phi,\quad  \frac{d\Phi}{dt}=B\Phi,
\quad
\Phi=
\left(\begin{array}{c}\Phi_1 \\\Phi_2\\\vdots\\ \Phi_N\end{array}\right),
\end{equation}
\begin{equation}
\fl
L=\left(
\begin{array}{ccccc}
\beta_1 & \alpha_1   &         &       & \\
\alpha_1 & \beta_2   & \alpha_2     &       & \\
    &\ddots & \ddots  &\ddots  & \\
    &       &  \alpha_{N-2}& \beta_{N-1}& \alpha_N\\
    &       &     0    & \alpha_{N-1}&\beta_N
\end{array}
\right),\quad
B=
\left(
\begin{array}{ccccc}
 0   & \alpha_1   &         &         & \\
-\alpha_1 & 0     & \alpha_2     &         & \\
     &\ddots & \ddots  &\ddots   & \\
     &       &- \alpha_{n-2}&  0      & \alpha_N\\
     &       &    0    &- \alpha_{N-1}&0
\end{array}
\right),
\end{equation}
is self-adjoint\cite{Flaschka}. The solutions of the linear
problem \eref{Toda:lin_I-V} and adjoint linear problem
\eref{Toda:adjlin_I-V} are related to $\Phi_n$ as
\begin{equation}
\Psi_n = {\rm e}^{-\mu t}(-1)^n{\rm e}^{-\frac{y_n}{2}}~\Phi_n,\quad
\Psi^*_n ={\rm e}^{\mu t} (-1)^n{\rm e}^{\frac{y_n}{2}}~\Phi_n,
\quad \mu=-\frac{1}{2}\lambda,\label{Flaschka_and_IV}
\end{equation}
respectively.
\begin{remark}
The relationship between entries of determinants and the
solutions of linear problems are given by applying Theorem
\ref{thm:Toda1} as
\begin{equation}
 \Omega^{(-1)}_0(t,\lambda)=\left[\frac{\Psi^*_1(t,\lambda)}{\Psi^*_0(t,\lambda)}\right]^{(-1)}
=\frac{1}{\lambda}\frac{1}{a_0^{(0)}}\sum_{i=0}^{\infty}a_i^{(0)}(-\lambda)^{-i}.
\label{gen_fn:semi-infinite}
\end{equation}
However it is not possible to express \eref{gen_fn:semi-infinite} in
terms of the solutions of the linear problem \eref{lin:Flaschka}
$\Phi_n$ by using the correspondence
\eref{Flaschka_and_IV}, since $\Phi_0$ is not defined for
the finite lattice. 
\end{remark}
\subsection{Results of Moser and Nakamura}
Moser \cite{Moser} considered  $(N,N)$ entry of the resolvent of matrix $L$:
\begin{equation}
f(\mu)=(\mu I-L)_{NN}^{-1}=\frac{\Delta_{N-1}}{\Delta_{N}},
\end{equation}
where $\Delta_n$ is given by
\begin{equation}
\Delta_n=
\left|
\begin{array}{ccccc}
\mu-\beta_{1} & -\alpha_{1}   &         &       & \\
-\alpha_{1} & \mu-\beta_{2}   & -\alpha_{2}     &       & \\
    &\ddots & \ddots  &\ddots  & \\
    &       &  -\alpha_{n-2}& \mu-\beta_{n-1}& -\alpha_{n-1}\\
    &       &     0    & -\alpha_{n-1}&\mu-\beta_n
\end{array}
\right|.\label{Delta} 
\end{equation}
We note that $f(\mu)$ is a rational function in $\mu$, since $\Delta_n$
is the $n$-th degree polynomial in $\mu$. By investigating analytic
properties of $f(\mu)$, Moser derived the action-angle variables of the
finite Toda equation to establish the complete integrability. Nakamura
\cite{Nakamura:Toda_tau} further investigated the expansion of $f(\mu)$
around $\mu=\infty$ to obtain
\begin{equation}
f(\mu)=\frac{\Delta_{N-1}}{\Delta_{N}}
=\frac{1}{\mu}\frac{1}{g_0}\sum_{i=0}^{\infty}g_i~(-2\mu)^{-i},\quad g_i'=g_{i+1},\label{Nakamura}
\end{equation}
and claimed that $g_i$ are the entries of the determinant formula
\eref{Darboux:2}, which is quite similar to our result. Let us discuss
this result from our point of view.

By expanding the determinant in \eref{Delta} with respect to $n$-th row,
we have the recurrence relation of $\Delta_n$
\begin{equation}
 \Delta_n=(\mu-\beta_n)\Delta_{n-1}-\alpha_{n-1}^2 \Delta_{n-2}.\label{rec_n:Delta}
\end{equation}
Also, one can show by induction
\begin{equation}
 \Delta_n' = -2\alpha_n^2 \Delta_{n-1}.\label{rec_t:Delta}
\end{equation}
Comparing \eref{rec_n:Delta} and \eref{rec_t:Delta} with the 
linear problems \eref{Toda:lin_I-V} and \eref{lin:Flaschka}, we have
from \eref{Toda:vars} and \eref{Flaschka_and_IV}
\begin{equation}
 \Delta_n = (-2)^{-n}~\Psi_{n+1} 
= 2^{-n}{\rm e}^{-\frac{y_n}{2}}~\Phi_{n+1} .\label{Delta_and_Psi}
\end{equation}
Now Proposition \ref{prop:det} and Theorem \ref{thm:Toda1} with $k=N$ yield
\begin{equation}
\begin{array}{c}
\medskip
{\displaystyle\frac{\tau_{N-n}}{\tau_N}=\det(b_{i+j-2}^{(N)})_{i,j=1,\ldots,n},}\\
\medskip
{\displaystyle b_i^{(N)} = b_{i-1}^{(N)\prime} + \frac{\tau_{N+1}}{\tau_N}\sum_{j=0}^{i-2}
  b_j^{(N)}b_{i-2-j}^{(N)}, \quad b_0^{(N)}=\frac{\tau_{N-1}}{\tau_N},}\\
{\displaystyle \left[\frac{\Psi_N(t,\lambda)}{\Psi_{N+1}(t,\lambda)}\right] ^{(-1)}=
\frac{1}{\lambda}\frac{\tau_N}{\tau_{N-1}}
\sum_{i=0}^\infty b_i^{(N)}~\lambda^{-i}.}
\end{array}
\label{tau_N}
\end{equation}
Then taking the boundary condition \eref{finite_Toda:bc2} into account,
noticing that $\Delta_n$ is polynomial of degree $n$ in $\mu=-\frac{\lambda}{2}$,
\Eref{tau_N} can be rewritten by using \eref{Delta_and_Psi} as
\begin{equation}
\begin{array}{c}
\medskip
 {\displaystyle 
\frac{\Delta_{N-1}}{\Delta_N}=
\frac{1}{\mu}\frac{\tau_N}{\tau_{N-1}}
\sum_{i=0}^\infty b_i^{(N)}~(-2\mu)^{-i},\quad
b_i^{(N)} = b_{i-1}^{(N)\prime},
}\\
{\displaystyle 
\frac{\tau_{N-n}}{\tau_N}=\det(b_{i+j-2}^{(N)})_{i,j=1,\ldots,n},
\quad b_0^{(N)}=\frac{\tau_{N-1}}{\tau_N},}
\end{array}
\end{equation}
which is nothing but (\ref{Nakamura}). In order to satisfy the boundary
condition (\ref{finite_Toda:bc1}) at the left edge of lattice, we choose
$b_0^{(N)}$ to be sum of $N$ terms of exponential function.

In summary, Nakamura's result may be interpreted as the determinant
formula viewed from opposite direction of the lattice. Namely, starting
from $n=N$ under normalization $\tau_N=1$, it describes such formula
that expresses $\tau_{N-n}$ in terms of $n\times n$ determinant. Since
$\tau$ function of the finite Toda equation is invariant with respect to
inversion of the lattice ($n\to N-n$), it is also possible to regard
this formula as expressing $\tau_n$ as $n\times n$ determinant under the
normalization $\tau_0=1$. Also, it should be remarked that the resolvent
of $L$ appeared because $\Delta_n$, principal minor determinant of 
$\mu I - L$, satisfies the auxiliary linear problem of the finite Toda
equation.

\begin{remark}
In order to obtain ``normal'' determinant formula,
we may consider $(1,1)$ entry of the resolvent of $L$
\begin{equation}
g(\mu)=(\mu I-L)_{11}^{-1}=\frac{\overline{\Delta}_{1}}{\overline{\Delta}_{0}},
\end{equation}
\begin{equation}
\overline{\Delta}_n=
\left|
\begin{array}{ccccc}
\mu-\beta_{n+1} & -\alpha_{n+1}   &         &       & \\
-\alpha_{n+1} & \mu-\beta_{n+2}   & -\alpha_{n+2}     &       & \\
    &\ddots & \ddots  &\ddots  & \\
    &       &  -\alpha_{N-2}& \mu-\beta_{N-1}& -\alpha_{N-1}\\
    &       &     0    & -\alpha_{N-1}&\mu-\beta_N
\end{array}
\right|.
\end{equation}
The recurrence relations for $\overline{\Delta}_n$ are given by
\begin{equation}
 \overline{\Delta}_n=(\mu-\beta_{n+1})\overline{\Delta}_{n+1}
-\alpha_{n+2}^2\overline{\Delta}_{n+2},\quad \overline{\Delta}_n' =
2\alpha_n^2 \overline{\Delta}_{n+1},
\label{lin:Deltabar}
\end{equation}
which implies
\begin{equation}
 \overline{\Delta}_n = (-2)^n~\Psi^*_n = 2^n{\rm e}^{\frac{y_n}{2}}~\Phi_n.
\end{equation}
Therefore Proposition \ref{prop:det} and Theorem \ref{thm:Toda1} yield
\begin{equation}
\begin{array}{c}
\medskip
{\displaystyle \frac{\overline{\Delta}_1}{\overline{\Delta}_0}=
\frac{1}{\mu}\frac{\tau_0}{\tau_{1}}
\sum_{i=0}^\infty a_i^{(0)}~(2\mu)^{-i},\quad a_i^{(0)} = a_{i-1}^{(0)\prime},
\quad a_0^{(0)}=\frac{\tau_{1}}{\tau_0},}\\
{\displaystyle 
\frac{\tau_{n}}{\tau_0}=\det(a_{i+j-2}^{(0)})_{i,j=1,\ldots,n}.}
\end{array}
\end{equation}
\end{remark}
\section{Concluding remarks}
In this article we have established the relationship between the Hankel
determinant formula and the auxiliary linear problem. We have also
presented a compact formula of the $\tau$ function in the framework of
the KP theory. The similar phenomena that have been observed in the
Painlev\'e II and IV equation can be recovered from this result. We have
also pointed out that Moser and Nakamura's result on the finite Toda
equation can be understood naturally in our framework. 

Since the Toda equation can be seen in various context, we expect that
the structure presented in this article can be observed in wide area of
physical and mathematical sciences. Moreover, it might be an intriguing
problem to study whether similar phenomenon can be observed or not for
the periodic lattice, where the theta functions play the role of the
$\tau$ functions.

\ack The authors would like to express their sincere thanks to N. Joshi
for fruitful discussions. They also thank Y. Nakamura and G. Carlet for
helpful conversation. This research was partially supported by MIMS, the
Manchester Institute for Mathematical Sciences. KK is supported by the
JSPS Grant-in-Aid for Scientific Research (B)15340057 and
(A)16204007. He also acknowledges the support by the 21st Century COE
program at the Faculty of Mathematics, Kyushu University.

\end{document}